\documentclass[aps,prb,twocolumn,groupedaddress]{revtex4}
\usepackage{graphicx}
\usepackage{epsfig}
\usepackage{float}
                                                                                
\begin{document}

\title{Electronic structure and bond competition in the polar magnet
PbVO$_3$}
\author{D.J. Singh}
\address{Condensed Matter Sciences Division, Oak Ridge
National Laboratory, Oak Ridge, TN 37831-6032, USA}

\begin{abstract}
Density functional electronic structure studies of tetragonal
PbVO$_3$ are reported. The results show a an important role
for both Pb 6$p$ - O 2$p$ and V $d$ - O $p$ bonding, with an interplay
between these. This is discussed in relation to the
possibility of obtaining magnetoelectric behavior.
\end{abstract}

\maketitle
\date{\today }

\section{Introduction}

Magnetoelectric and specifically multiferroic magnetoelectrics have
attracted renewed attention, \cite{spaldin,fiebig,hill,kimura}
although the field has a long history.
\cite{curie,smolenskii,smol2,smol3,smol4}
Nonetheless, bulk materials that provide strong coupling between electric and
magnetic fields at room temperature remain elusive. Some of the difficulties
in realizing such materials were reviewed by Hill. \cite{hill}
Perovskite and perovskite derived phases, $AB$O$_3$,
provide the main inorganic
ferroelectric materials used in applications.
These are based on a cubic lattice with $A$ ions on the cube corners and
$B$ ions
at the cube centers,
inside corner sharing octahedral O cages,
Ferroelectric perovskites can be classified as $A$-site or $B$-site driven
according to which ion is primarily responsible for the polar instability.
One strategy for synthesizing multiferroics is to substitute
one of the perovskite sites with magnetic ions in materials where
the ferroelectricity is driven by the other site, as for example
in BiFeO$_3$. \cite{ramesh}
However, in both $A$ and $B$-site
driven materials the ferroelectricity generally involves cooperation
between off-centerings of both cations, and furthermore covalency between
unoccupied metal states and O $p$ states is important for the
ferroelectric instability. \cite{hk,cohen,llb,dskt,grinberg,ghita}
From this point of view, V$^{4+}$ may be a particularly interesting ion
for multiferroics.

In oxides, V$^{4+}$ is often
magnetic and, significantly, can occur either at the centers of
octahedral cages, as in CaVO$_3$ and SrVO$_3$, in which case there
are interesting metal insulator transitions of Mott character,
\cite{fujimori,inoue}
or it can strongly bind to one O forming forming a vanadyl ion with
very short bond length,
$\sim 1.55-1.60$ \AA. These vanadyl containing compouds are
local moment magnets due to the formation of spin 1/2 moments on the V ions,
and are band insulators, The band gaps are formed by a combination
of strong crystal field splitting associated
with the short V-O distance, and on-site Hund's coupling, which exchange
splits the 1/2 filled V non-bonding orbital. \cite{pickett,rai}
Thus, perovskite vanadates offer the possibility of combining
a magnetic $B$-site ion with insulating behavior and stereochemical
activity on both the $A$ and $B$-sites. This was recently realized
by the high pressure synthesis of PbVO$_3$. \cite{shpanchenko,belik}

PbVO$_3$ is insulating and
occurs in the tetragonal, non-centrosymmetric structure,
spacegroup $P4/mm$, like PbTiO$_3$, but with a very large tetragonality,
$c/a \sim 1.23$. The structural distortion may be described as off-centering
of Pb and V accompanied by a distortion of the VO$_6$ octahedral cage. 
This is in contrast to CaVO$_3$ and SrVO$_3$, which have centrosymmetric
GdFeO$_3$ type structures deriving from octahedral tilting.
PbVO$_3$ shows no structural transitions from 0K to
the decomposition temperature. However, a tetragonal to cubic
transition does occur under pressure, \cite{belik}
indicating the possibility
of ferroelectric behavior under pressure and
perhaps by chemical modification. It is noteworthy
that at 300K the shortest V - O distance is 1.67 \AA, which is
longer than in most other vanadyl containing oxides.
This implies that the vanadyl bond is relatively weak in PbVO$_3$.
Furthermore, because of the connection between the 
short V - O bond and the moment formation, one may anticipate
the possibility of coupling between magnetism and the structural
distortion, which in turn, if ferroelectric, would be coupled to external
electric field. Significantly, the Pb displacement with respect
to the center of its O cage is 0.92 \AA, and the V displacement
is 0.60 \AA. These are large and of comparable magnitude,
indicating both robust ferroelectricity if
switching could be achieved, and cooperativity between the two cation sites,
which is a characteristic of good ferroelectric perovskites.

Shpanchenko and co-workers \cite{shpanchenko} reported local spin density
approximation (LSDA) electronic structure calculations and an analysis
of the bonding and magnetism. These calculations were based on the
linearized muffin-tin orbital method with the atomic spheres
approximation (LMTO-ASA). This method is well suited to magnetic
transition metal oxides, but is suffers from shape and basis set
errors in open structures with highly non-cubic site symmetries, as
is the case for PbVO$_3$.
They also reported LSDA+U calculations for ferromagnetic ordering only,
with the augmented planewave plus local orbital (APW+LO) method.
The comparison of these two leads to the conclusion that the formation
of the spin polarized vanadyl bond depends on Hubbard
correlation effects.
Here we report LDA calculations done with
the general potential linearized augmented planewave (LAPW) method, which
incorporates the full crystal potential and has a much better basis
set than the LMTO-ASA method, especially for the interstitial regions.
\cite{singh-book}
We use these to discuss the formation of the electronic structure
and the bonding of the compound.
We find results that differ significantly from those of Ref.
\onlinecite{shpanchenko},
with a different conclusion regarding the vanadyl bond.
Our results compare well with very recent calculations of Uratani
and co-workers. \cite{uratani}
We find an interplay between Pb-O and V-O covalency that
may be of importance for magnetoelectric materials.

\section{Computational Method}

As mentioned, the calculations were done using the general potential LAPW
method. Except as noted, the experimental 300K crystal structure reported
by Belik and co-workers \cite{belik} was used.
The calculated LSDA forces on the atoms are small
with this structure,
supporting the experimental determination as well as the
conclusion that the vanadyl bond is properly formed at the
LSDA level. The
largest force is on the apical O of 0.032 Ry/$a_0$ towards the V, with
a smaller force of 0.017 Ry/$a_0$ drawing the V towards the apical O for
the lowest energy (see below) antiferromagnetic ordering.
The LAPW sphere radii were 2.00 $a_0$, 1.70 $a_0$, and 1.35 $a_0$, 
respectively for Pb, V and O. Local orbitals \cite{singh-lo}
were used to treat the
high lying valence states of Pb and V, as well as to relax linearization
errors. The basis sets consisted of approximately 1100 LAPW functions plus
local orbitals for the primitive cells and correspondingly larger sizes
for supercells. Zone sampling during the iterations to self-consistency
was done using the special points method, with 40 {\bf k}-points
in the irreducible wedge for the primitive cell. Tests with other
choices were made. These showed that the calculations were well converged.
Magnetic ordering was studied using $\sqrt{2} \times \sqrt{2} \times 2$
supercells. In all cases, the core states were treated self-consistently
and fully relativisically,
while the valence states were calculated
in a scalar relativistic approximation.

\section{Moment Formation}

Stable spin moments of 1 $\mu_B$ were found for all magnetic configurations,
consistent with local moment magnetism.
This differs from the calculations reported in Ref. \onlinecite{shpanchenko},
where the F ordering showed only a small moment in contrast to the C ordering.
As mentioned, they also did LSDA+U calculations for the
ferromagnetic case, but with the
APW+LO method instead of the LMTO-ASA method that they used for the other
calculations in their paper. The implication of their results is that
the Hubbard U is essential for a stable spin polarized vanadyl bond.
We find a stable spin polarized bond in accurate LSDA calculations, 
with the implication that the key physics is hybridization with O
and on-site Hund's coupling.
Calculations were done for
an artificial non-spin-polarized case, as well as ferromagnetic (F),
antiferromagnetic in the $ab$ plane (C), antiferromagnetic along all
three directions (G), and ferromagnetic $ab$ planes stacked
antiferromagnetically along the $c$-direction (A). The calculated energies
are given in Table \ref{tab-e}.

\begin{table}[tbp]
\caption{Energetics of various spin configurations of
PbVO$_3$ relative to the non-spin-polarized case. Energies are
given in meV per formula unit.}
\begin{tabular}{cccc}
\\
\hline
 F & G & A & C \\
\hline
 ~~~~-111.2~~~~ & ~~~~-127.3~~~~ & ~~~~-91.8~~~~ & ~~~~-127.8~~~~ \\
\hline
\end{tabular}
\label{tab-e}
\end{table}

We find the
lowest energy state to be antiferromagnetic type C, qualitatively in
agreement with
the calculations of Ref. \onlinecite{shpanchenko}, and with energies
in good agreement with those of Ref. \onlinecite{uratani}.
We note that
the energies are not well described by a nearest neighbor
Heisenberg model. Both the C and G orderings have antiferromagnetic
$ab$ planes but opposite stacking along $c$. Similarly, F and A
have ferromagnetic $ab$ planes, again differing in the $c$ stacking.
However, the energies of C and G differ by less than 1 meV per
V, while those of F and A differ by approximately 19 meV. Thus interactions
at least to second neighbor are needed to reconcile the
energies. Considering this,
as well as the extremely small energy difference between the C and G
orderings, the actual ground state could be complex.
One may speculate
that this is the reason for non-observation of antiferromagnetic
superlattice peaks in neutron diffraction.
We note that there is 
evidence for a transition at $\sim$120K seen both in thermal expansion
\cite{belik}
and resistivity. \cite{shpanchenko}.

\section{Band Structure}

\begin{figure}[tbp]
\vskip 0.3cm
\epsfig{width=0.65\columnwidth,angle=270,file=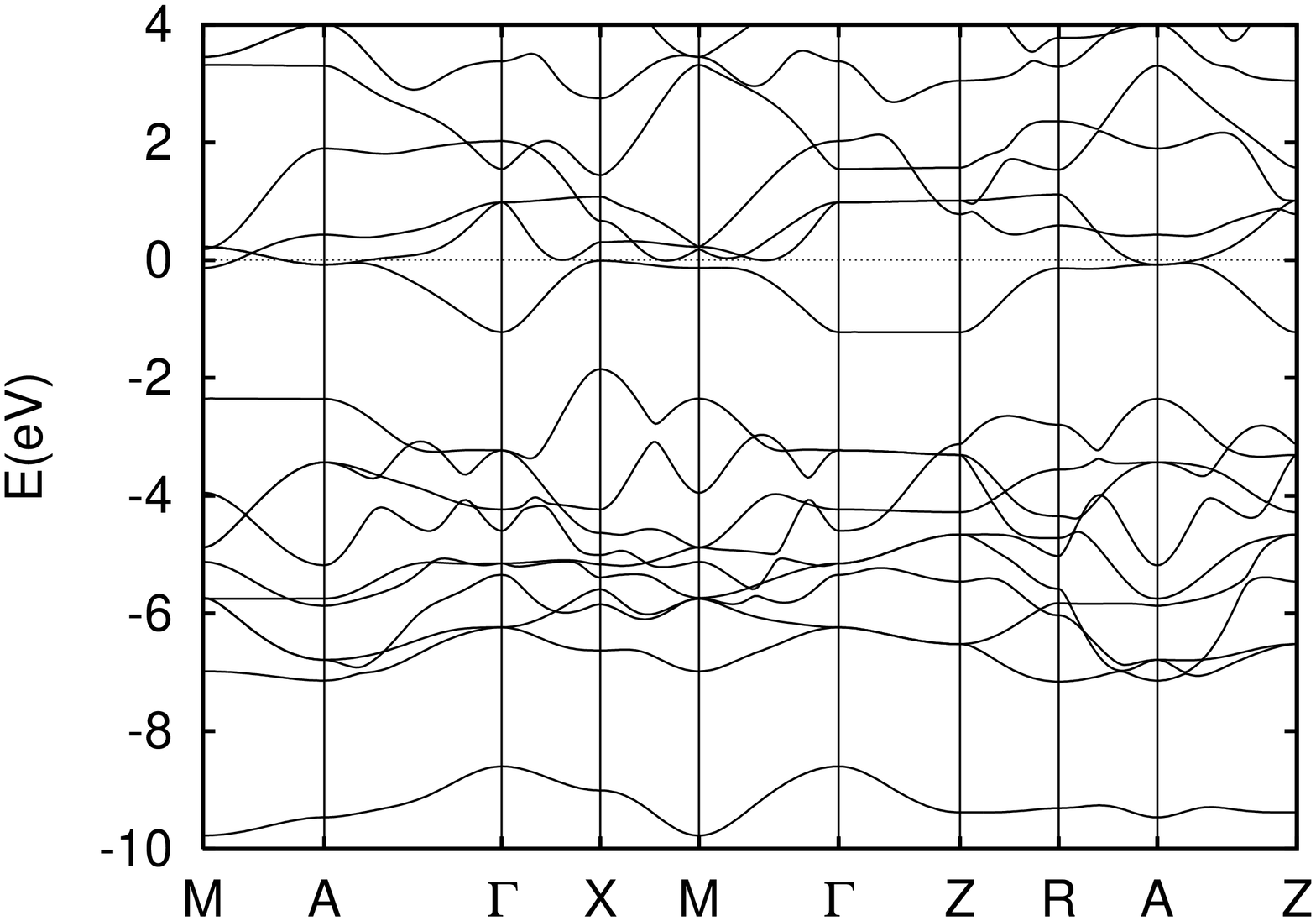}
\vskip 0.3cm
\epsfig{width=0.65\columnwidth,angle=270,file=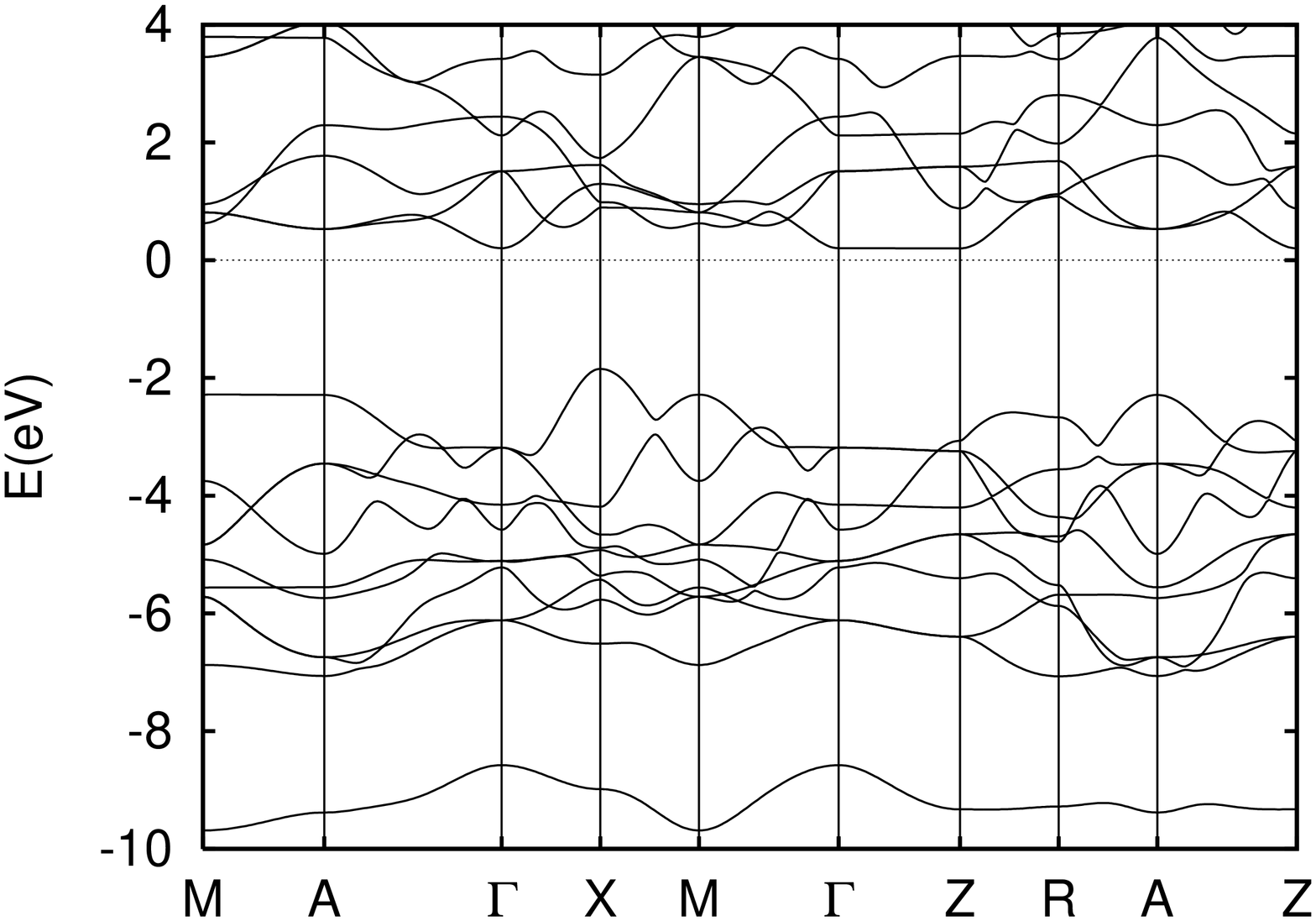}
\vskip 0.3cm
\caption{Band structure of ferromagnetic ordered PbVO$_3$. The
top panel shows the majority spin bands and the bottom shows the
minority spin.}
\label{bands-f}
\end{figure}

\begin{figure}[tbp]
\vskip 0.3cm
\epsfig{width=0.65\columnwidth,angle=270,file=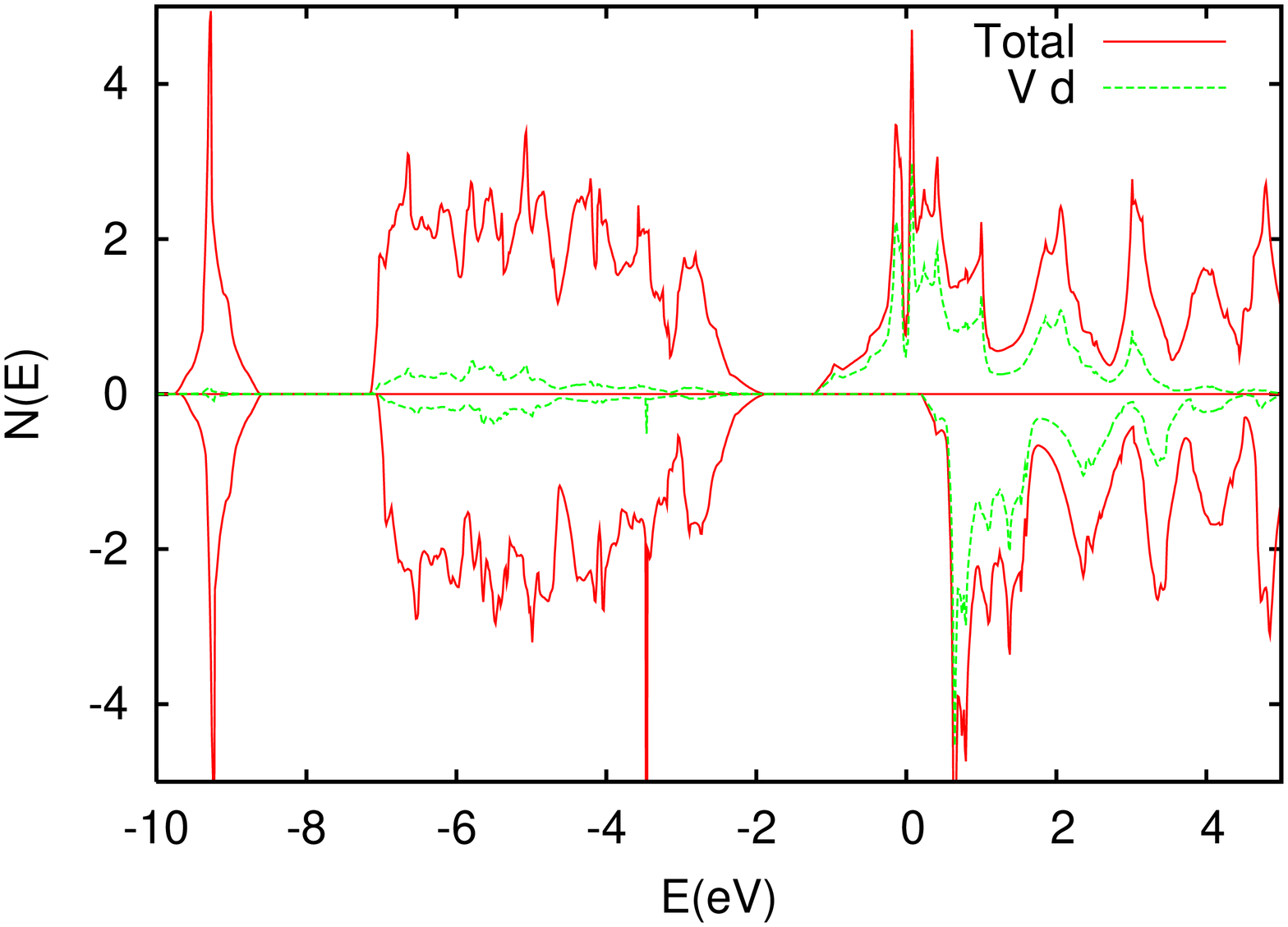}
\vskip 0.3cm
\epsfig{width=0.65\columnwidth,angle=270,file=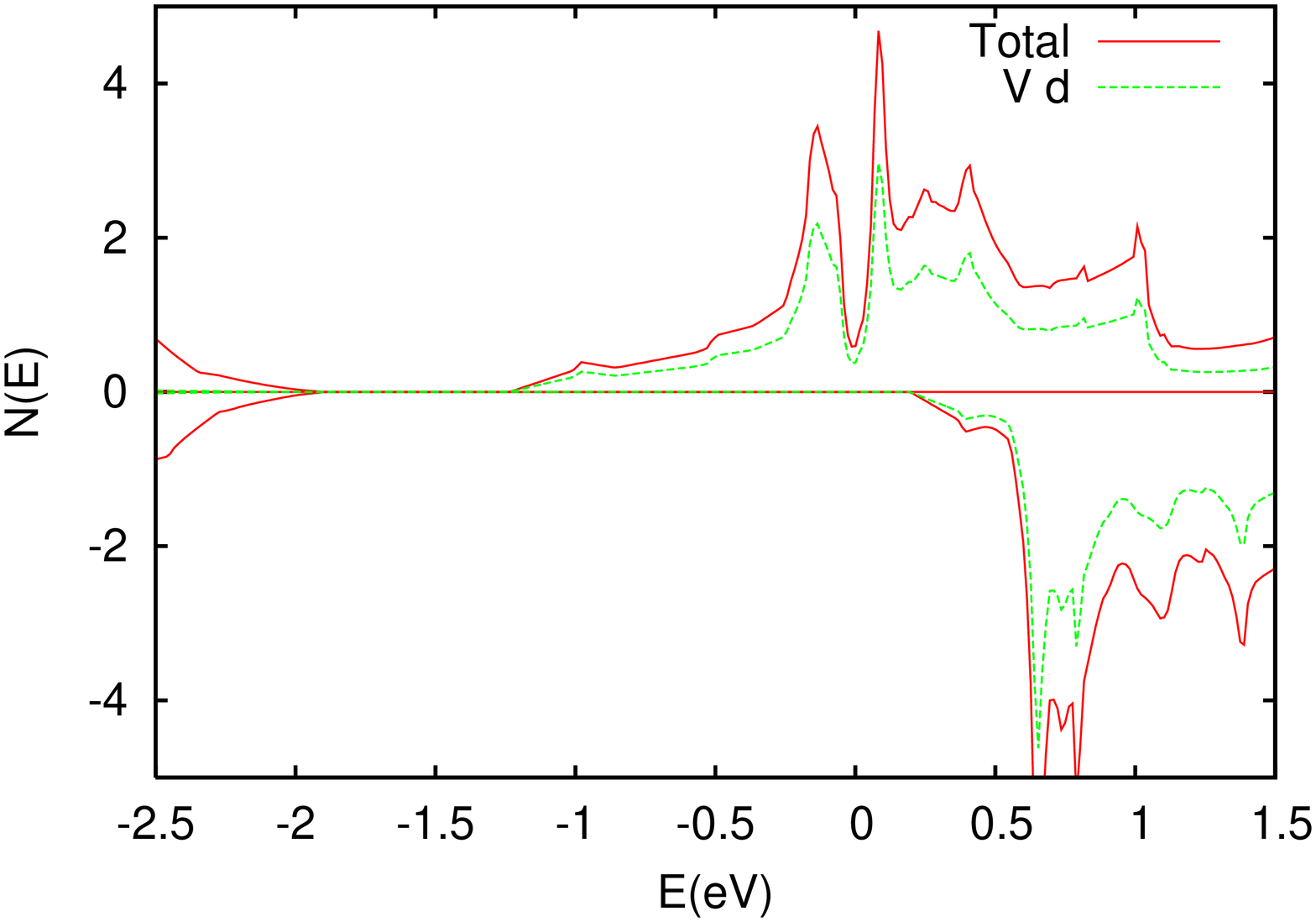}
\vskip 0.3cm
\epsfig{width=0.65\columnwidth,angle=270,file=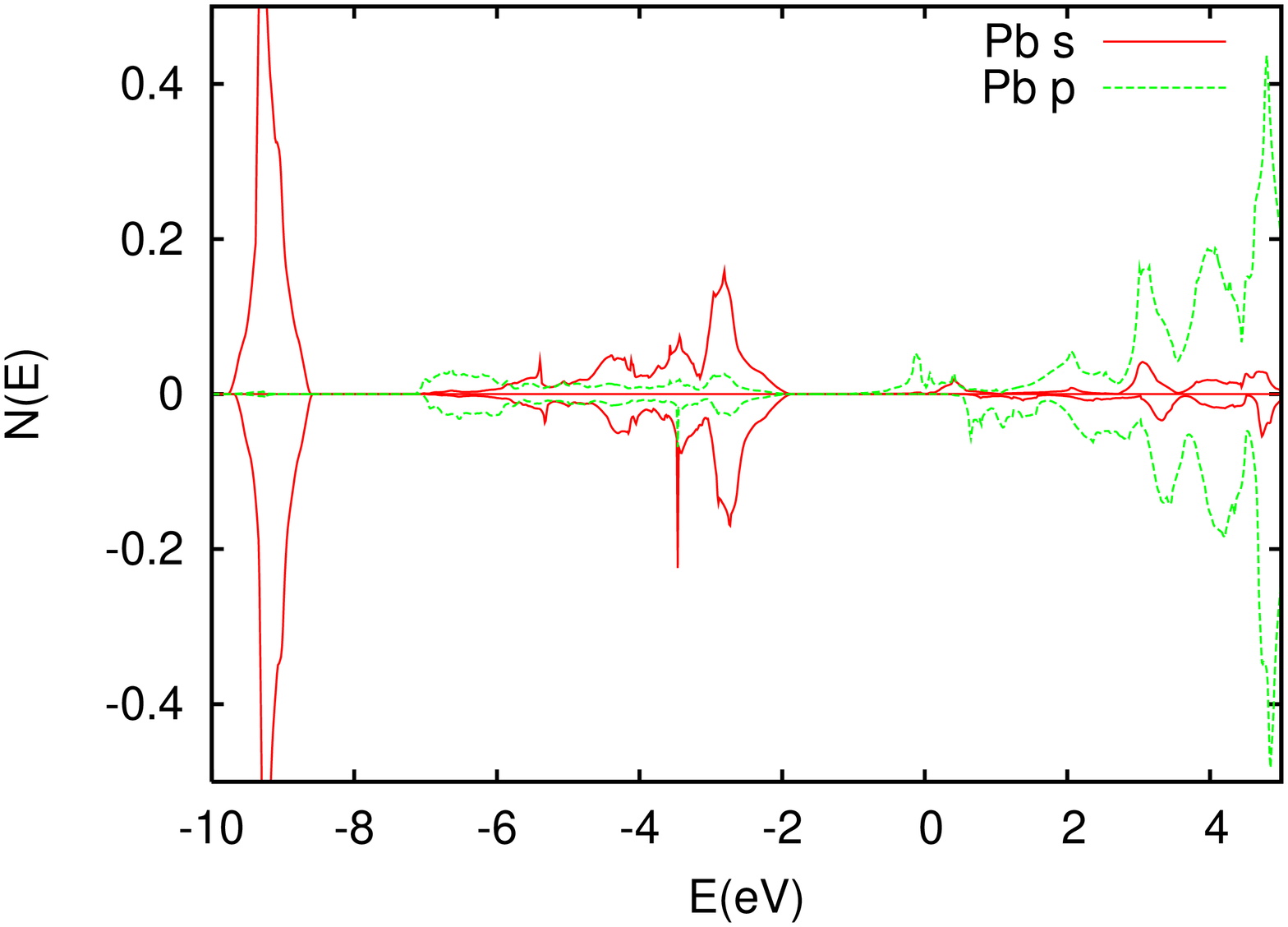}
\vskip 0.3cm
\caption{(color online) Electronic DOS and projections
onto LAPW spheres for ferromagnetic PbVO$_3$. The
top panel shows the total DOS and projections onto the
V $d$ character, while the bottom shows Pb contributions.
The middle panel is a blow up around $E_F$.
Note that since the Pb orbitals are more extended than the
2$a_0$ sphere radii, the Pb projection is proportional to the
corresponding contributions, but reduced in magnitude. Majority
and minority spin are shown above and below the axes, respectively.}
\label{dos-f}
\end{figure}

We now turn to the electronic structure, starting with the F ordering.
The band structure is shown in Fig. \ref{bands-f}, and the corresponding
density of states (DOS) and projections in Fig. \ref{dos-f}.
The band structure, in order of increasing energy,
consists of a split-off Pb $6s$ derived peak, approximately
9 eV below the Fermi energy followed by a manifold of O $2p$ derived bands,
and then, following another gap, a manifold of metal derived bands. Initially,
these are V $d$ derived, but at higher energy ($\sim$ 3 eV) these change
to include Pb $p$ and other characters.

\begin{figure}[tbp]
\vskip 0.3cm
\epsfig{width=0.65\columnwidth,angle=270,file=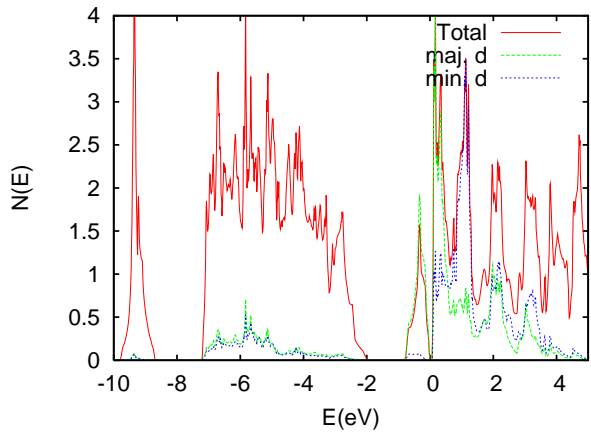}
\vskip 0.3cm
\caption{(color online) Electronic DOS of PbVO$_3$ for the lowest
energy C-type antiferromagnetic ordering.}
\label{dos-qc}
\end{figure}

The crystal field may be regarded as
a combination of an octahedral crystal field, which splits the
$d$ levels into a lower lying $t_{2g}$ group and a higher lying
$e_g$ group. These are then further split by the off-centering.
The main contribution to the crystal field in oxides is from metal -
oxygen hybridization. Since the O 2$p$ bands are below the
V $d$ bands, the V $d$ orbitals are formally anti-bonding, so the
the stronger the hybridization of a given $d$ orbital with O, the
higher it will be pushed. The result is a splitting of the $e_g$
manifold into two and a splitting of the $t_{2g}$ manifold into
a low lying $d_{xy}$ non-bonding band and a higher two-fold degenerate
$d_{xz}$ and $d_{yz}$ derived manifold. This explains the multi-peak
structure of the $d$ band DOS. In the case of V$^{4+}$, there is
one $d$ electron in the $d_{xy}$ orbital, which forms local
moments due to the Hund's coupling. The exchange splitting for F ordering
is 1.43 eV, and not sensitive to the magnetic ordering (see
Fig. \ref{dos-qc}).
This is sufficient to fully polarize the $d$ orbitals.
For F ordering this yields
a half-metallic state, spin moment 1 $\mu_B$.

With the experimental structure of
PbVO$_3$ and F ordering,
the crystal field splitting is almost, but not quite, sufficient
to open a complete gap between the majority spin $d_{xy}$ band
and the higher lying $d$ bands. This is because of the bandwidth of
the $d_{xy}$ band. As may be seen in majority spin panel of Fig. \ref{bands-f},
the $d_{xy}$ band has almost no dispersion along the $\Gamma-Z$ line and
has weak dispersion along $M-A$ implying that the main hopping is the
$ab$ plane. Following the dispersion along $\Gamma-M$, the $d_{xy}$ band maximum
is 0.3 eV above the bottom of the other $d$ bands.

We did a structural relaxation for the F ordering. The resulting LSDA
positional parameters are
$z_{\rm V}$=0.5689, $z_{\rm O1}$=0.2158, and $z_{\rm O2}$=0.6872,
yielding a V - O1 bond length of 1.67 \AA, which is slightly less than
the experimental value. The
corresponding experimental values are
$z_{\rm V}$=0.5668, $z_{\rm O1}$=0.2102, and $z_{\rm O2}$=0.6889.
The calculated full symmetry $a_g$ Raman frequencies are
190 cm$^{-1}$ (Pb motion against the other atoms in the cell),
408 cm$^{-1}$ (cations vs. O),
and 838 cm$^{-3}$ (V - O1 bond stretching, reflecting the short bond length
and stiff interaction).

The effect of in-plane antiferromagnetism is to narrow the $d_{xy}$
band by disrupting nearest neighbor hopping. This is because the
exchange spitting is large. The result is a band gap as shown in Fig.
\ref{dos-qc}. Because the crystal field splitting is sensitive to
the exact V - apical O distance, the band gap should also be sensitive
to this structural parameter. We can also understand in these terms why
the energy difference between the G and C type orderings is small
and that between the F and A type orderings is not, even though
both pairs differ only in the $c$ direction magnetic stacking. Specifically,
in the
F and A orderings, the majority $d_{xy}$ manifold overlaps the other $d$
states so that the moment is not of pure $d_{xy}$ character, and there
are non-trivial dispersions in the $c$ direction in the region of the zone
where there is overlap,
such as $M-A$ of Fig. \ref{bands-f}. These dispersions
provide interaction along
$c$, and because they consist of
is metallic $d$ band overlap in only one spin channel,
the resulting spin
interaction is strongly ferromagnetic. Thus F is strongly favored over A.
On the other hand, with antiferromagnetic in-plane ordering, the $d$
bands are narrower, and so the
$d_{xy}$ band does not overlap the other $d$ states. Thus the
$c$ axis exchange coupling is much weaker. At the experimental structure
it is very weakly ferromagnetic, favoring C over G, but if the gap were
larger, either due to a shortening of the V - apical O distance or
on-site Coulomb repulsions, one would expect it to become antiferromagnetic.
In this case, if the magnetic structure is simple, the ground state would
become G type. Thus, at least at the LSDA level, PbVO$_3$ is near a
structure dependent cross-over between two magnetic states.

\section{Role of Pb}

Turning now to the role of Pb, the bottom panel of Fig. \ref{dos-f} shows
projections of the DOS onto the Pb LAPW sphere (radius 2 $a_0$). The
large peak below the O $2p$ bands is from the Pb $6s$ states. As may be
seen, these mix with the occupied
O $2p$ bands, but play almost no role in the unoccupied conduction
bands. Mixing of occupied states does not produce bonding.
On the other hand, the Pb $6p$ states are low lying in the conduction
bands and mix with the O $2p$ bands, which are the same bands
that hybridize with V $d$ states.
Thus there is a mechanism for competition between
the vanadyl bonding and the Pb - O bonding.

We emphasize that while the lone 6s pair of Pb$^{2+}$ is often
discussed in terms of the stereochemical activity of Pb, these orbitals
are not responsible for the activity directly. Rather it is that
in Tl(I), Pb(II) and Bi(III) the atomic structure that leads to the
position of the 6s orbital also leads to 6p levels that are
low lying in the conduction bands and very spatially extended favoring
hybridization with occupied ligand orbitals.
We emphasize that while the lone 6s pair of Pb$^{2+}$ is often
discussed in terms of the stereochemical activity of Pb, these orbitals
are not responsible for this activity directly. Rather it is that
in Pb(II) and Bi(III) the atomic structure that leads to the
position of the 6s orbital also leads to 6p levels that are
low lying in the conduction bands and very spatially extended favoring
hybridization with occupied ligand orbitals.

\begin{figure}[tbp]
\vskip 0.3cm
\epsfig{width=0.65\columnwidth,angle=270,file=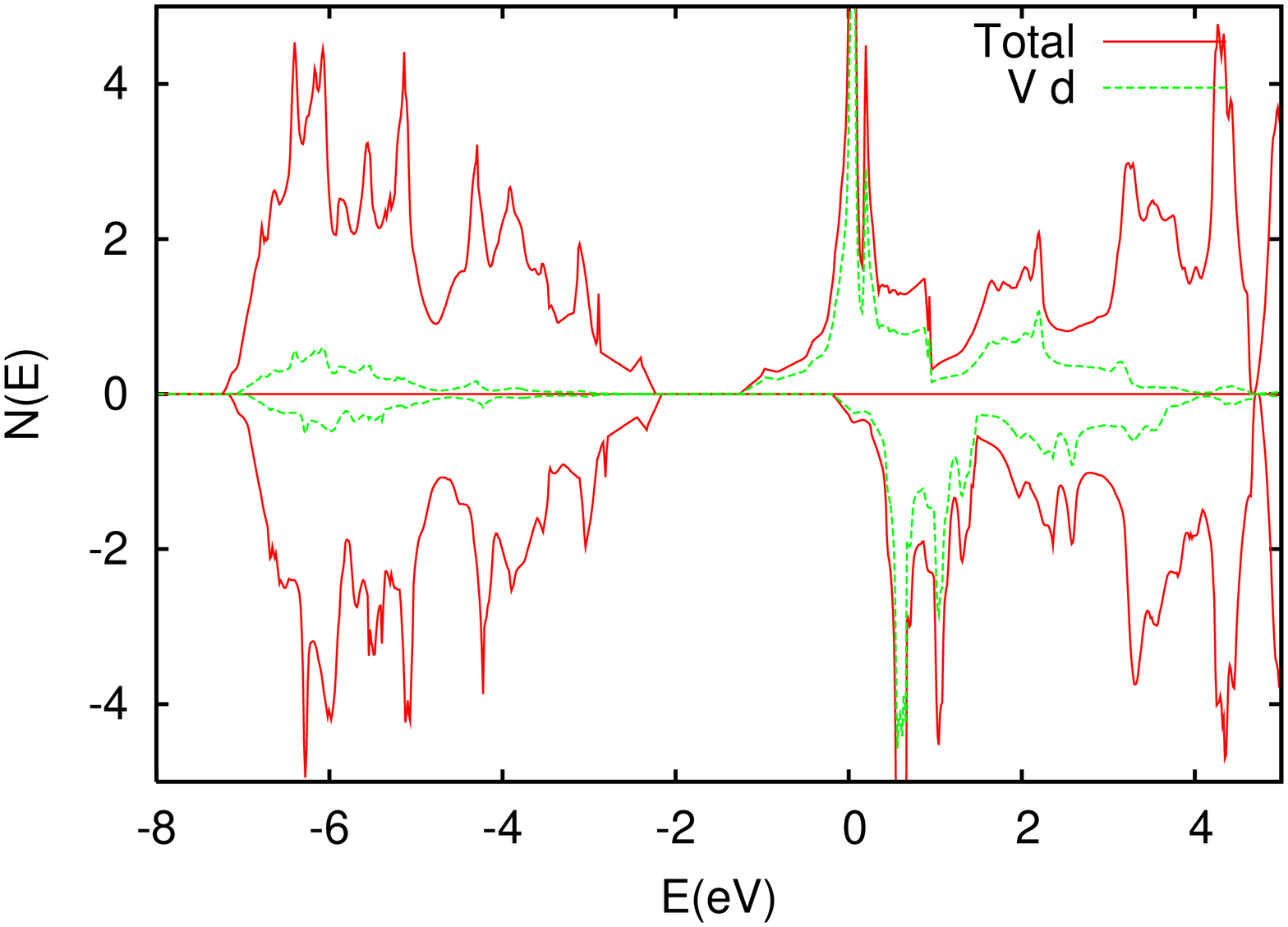}
\vskip 0.3cm
\epsfig{width=0.65\columnwidth,angle=270,file=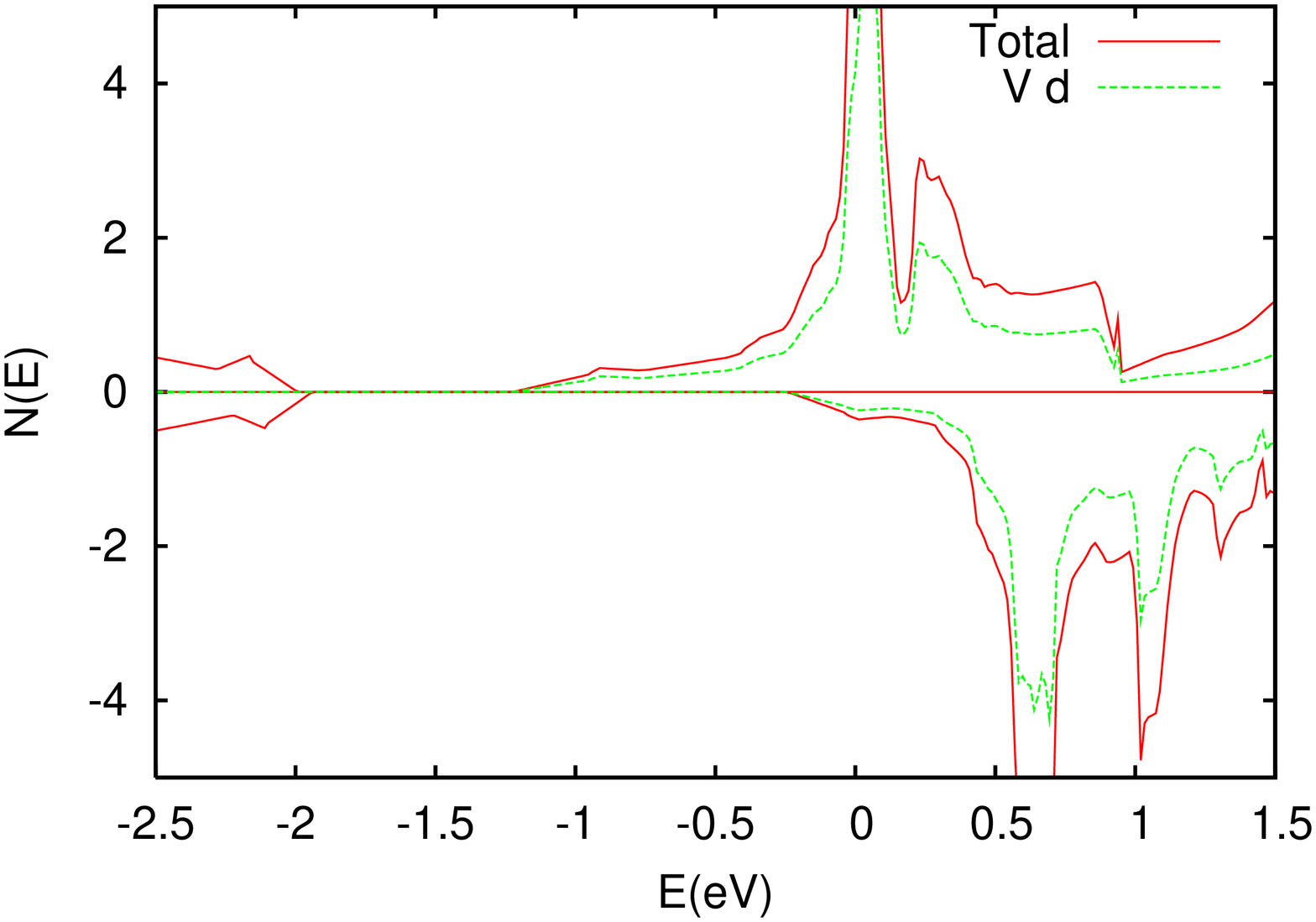}
\vskip 0.3cm
\caption{(color online) Electronic DOS of hypothetical ferromagnetic CaVO$_3$
constrained to the PbVO$_3$ crystal structure (see text) and a blow-up
around $E_F$.
}
\label{dos-cf}
\end{figure}

Sr$^{2+}$ is similar in size to Pb$^{2+}$,
but does not have the lone pair and the
resulting low lying spatially extended unoccupied $p$ states. As a
result, it would not form structures with short bond lengths to O
like the reported structure of PbVO$_3$. Ca$^{2+}$ is a smaller ion,
also lacking the lone pair,
and is more compatible with the bond lengths in PbVO$_3$, though, as
mentioned, CaVO$_3$ actually forms a centrosymmetric
tilted GdFeO$_3$ type structure.
As such, we did calculations for hypothetical
ferromagnetic ordered CaVO$_3$
with the atomic positions fixed at those of PbVO$_3$
in order to elucidate the role of Pb.
The calculated DOS is shown in Fig. \ref{dos-cf}. As may be seen,
the pseudogap between the majority spin $d_{xy}$ band and the remaining
$d$ bands is reduced and the half-metallic character is lost (the
spin moment is 0.88 $\mu_B$ instead of 1 $\mu_B$). As mentioned,
in PbVO$_3$ the calculated forces on the atoms are small. Keeping the
structure fixed, there is a strong force of 0.08 Ry/$a_0$ pushing the
V away from the apical O, i.e. disrupting the vanadyl bond.
Thus indeed there is substantial coupling between the Pb-O
interaction and the V-O interaction, seen both in the stability
of the vanadyl bond and the magnetic properties.
This is due to bond competition with O. Specifically,
the removal of Pb leads to broadening of the $t_{2g}$ bands, which
which spoils the crystal field scheme that stabilizes the
vanadyl bond.

\section{Summary}

To summarize, LSDA calculations support the experimental conjecture
that PbVO$_3$ may be the basis for interesting magnetoelectric
materials. This is based on the electronic coupling between
Pb and V due to mutual bonding with O and the
stabilization of magnetism by a mechanism connected with
the V - O bonding. The key challenge is
to find chemical modifications that make the material more
switchable, so that ferroelectricity can be observed.

\acknowledgements

We are grateful for helpful discussions with Y. Shimakawa.
This work was supported by the U.S. Department of Energy and the
Office of Naval Research.

\end{document}